\documentclass[aps,pre,amsmath,twocolumn,amssymb,floatfixng,showpacs,longbibliography,
superscriptaddress,nofootinbib]{revtex4-2}
\usepackage{amsfonts}
\usepackage{amsmath}
\usepackage{amssymb}
\usepackage{latexsym}
\usepackage{array}
\newcolumntype{P}[1]{>{\centering\arraybackslash}p{#1}}
\newcolumntype{M}[1]{>{\centering\arraybackslash}m{#1}}
\usepackage{caption}
\usepackage[colorlinks,citecolor=red,urlcolor=blue,bookmarks=false,hypertexnames=true]{hyperref} 
\usepackage{xcolor}
\usepackage{lipsum}
\usepackage{bm}
\usepackage{relsize}
\usepackage{float}
\usepackage{subfloat}
\usepackage{dcolumn}
\usepackage{epsfig}
\usepackage{graphicx}
\usepackage{multirow}
\usepackage{makecell}
\usepackage{mathtools}
\usepackage[bbgreekl]{mathbbol}

\usepackage{footnote}
\newcommand\ra{\rangle}

\begin{document}

\title{Quantum Langevin Equation of a spin in a magnetic field : an analysis}
\author{Suraka Bhattacharjee}
 \affiliation{Raman Research Institute, Bangalore-560080, India} 
\author{Koushik Mandal}
\thanks{Present address: Institute of Physics, Sachivalaya Marg, Bhubaneswar-751005, India}
\affiliation{S. N. Bose National Centre for Basic Sciences, Kolkata-700106, India}

 \author{Supurna Sinha}
 \affiliation{Raman Research Institute, Bangalore-560080, India}

\date{\today}

\begin{abstract}
We derive a quantum  Langevin equation for a quantum spin in the presence of a magnetic field and study its dynamics in the Markovian limit using the Ohmic bath model. We extend our analysis to the Drude bath with a finite memory.
We study the time evolution of the expectation values of the magnetic moments. 
The spin auto-correlation functions exhibit a damped oscillatory behaviour with the randomization time being determined by 
the damping rate and also the memory time for the Drude bath model. We also analyse the spin response function of the system for the Ohmic bath model. 
Our results are consistent with findings in cold atom experiments. In addition we make predictions which can be tested in future ultra cold atom experiments.
\end{abstract}

\maketitle
\section{Introduction}
 \vspace{0.3cm}
Spin relaxation dynamics is central to many condensed matter systems \cite{Gilbert,abragam1961principles,slichter1963principles,deutch1968time,oppenheim1973electron,Cukier,fluctuation1962relaxation}. 
In recent years the emergence of nanoscale quantum technology involving spin systems, 
for instance spin based quantum computers \cite{nature}, has led to a renewed interest 
in understanding the dynamics of dissipation and damping due to contact with 
the environment. It is therefore important to understand the dissipative
dynamics of a spin in contact with an environment in the quantum domain. 
This is one of the motivations behind the present study. 

The motion of a classical Brownian spin interacting with particles in a medium has been studied fairly extensively \cite{Kubo1,Kawabata}. A theoretical analysis of such a system involves a stochastic model \cite{Seshadri1982}
where one replaces the effect of the surrounding medium by a random 
time dependent magnetic field. 

Before we discuss our work we briefly review earlier related studies. 
In \cite{Jayanavar1991} Langevin dynamical equations have been 
derived for the various spin components for a spin in a magnetic field.
These equations are valid in the high temperature classical domain. 
Later, the work of Ref.\cite{Jayanavar1991} was extended to include both linear and linear-quadratic couplings between the particle and the heat bath variables \cite{Palacios2}. Several other approaches for exploring the spin dynamics in the classical domain include the study of a classical spin coupled to a heat bath via a phenomenological Langevin equation and associated Fokker-Planck equation for the spin \cite{Miyazaki}. In \cite{Dudarev2011} the ensemble averages of the spin components and the statistical moments for $N$ coupled spins have been derived using a Langevin dynamics in the framework of a semi-classical perturbative approach. Langevin spin dynamics based on an ab initio calculation and other numerical schemes have been applied to calculate the magnetic moments, starting from the stochastic Landau-Lifshitz-Gilbert equation for a one dimensional Heisenberg chain \cite{Rozsa}. Moreover, the Landau-Lifshitz-Gilbert equation for small magnetic particles has been solved numerically using Stratonovich stochastic calculus and the trajectories of the individual magnetic moments have been studied via a Langevin dynamics approach \cite{Palacios}.
A generalized Langevin spin dynamics algorithm has been developed in \cite{Duradev3} where both the longitudinal and transverse degrees of freedom of atomic magnetic moments are treated as dynamic variables. 
Furthermore, molecular dynamics simulations were used to determine the temperature of a dynamical spin ensemble, expressed in terms of dynamic spin variables, 
by solving the semiclassical Langevin spin-dynamics equations and applying the fluctuation-dissipation theorem \cite{Duradev2}. The Langevin formalism has been further used to study the ultrafast spin dynamics for interacting spins coupled to a non-Markovian heat bath \cite{Atxitia}. A path integral approach has been used to study the spin dynamics equations using a generalized coherent state representation for the bosonic operators pertaining to the bath \cite{Rebei}. 
In a recent study \cite{Anders}, the authors have made use of the Landau-Lifshitz-Gilbert equation to develop a Heisenberg-Langevin equation for exploring the dynamics of a quantum spin coupled to an environment. The study has been extended to a semi-classical numerical 
simulation of a non-Markovian Lorentzian bath and its Ohmic limit \cite{Anders}.

In this paper we go beyond these analyses and derive quantum 
Langevin equations for the magnetic moments of a quantum spin coupled to a heat bath consisting of a large number of harmonic oscillators and use a rigorous quantum mechanical non-perturbative scheme to analytically
solve the equations and analyse the dynamics of the expectation 
values of the magnetic moments, the spin auto-correlation functions and the 
spin response function. The results agree with that of the classical and semi-classical 
analyses when one considers $\hbar \rightarrow 0$ \cite{Jayanavar1991}.
The predictions of our analysis can be tested against cold atom 
experiments. 

The paper is organised as follows. In Sec. $II$ we derive the quantum 
Langevin equations for a spin in a magnetic field and in contact with 
a heat bath. We consider two baths - the Ohmic bath and the Drude bath.
We evaluate analytical expressions for the expectation values of the magnetic moments and the spin 
auto-correlation functions for the two cases. In addition, we analytically evaluate the 
spin response function for the Ohmic bath model. 
In Sec. $III$ we display the plots that stem from our analysis and 
discuss our results. We finally end the paper with a few concluding 
remarks in Sec. $IV$. We display some of the detailed calculations in 
an Appendix.

\section{Quantum Langevin formalism for a Spin in a magnetic field}
\vspace{0.3cm}
We study the quantum Langevin dynamics of a Brownian spin in the presence of a magnetic field and in contact with a heat-bath. 
Our starting point is the  Hamiltonian for a spin coupled to the heat bath and in an external magnetic field :  
\begin{equation}
    H = -M_{z}H_{0} +\sum_{j=1}^3\sum_{\alpha} \frac{p_{\alpha j}^{2}}{2m_{\alpha j}}+\frac{1}{2}m_{\alpha j}\omega_{\alpha j}^{2}\left(q_{\alpha j}+\frac{\lambda_{\alpha j}M_{j}}{m_{\alpha j}\omega_{\alpha j}^{2}}\right)^{2}
    \label{spinlangevin}
\end{equation}
where, $H_{0}$ is the external magnetic field applied along the z-axis, $M_{j}(j=x,y,z)$ is the spin magnetic moment in the j$^{th}$ direction and $p_{\alpha j}$,  $q_{\alpha j}$, $\omega_{\alpha j}$, $m_{\alpha j}$ are respectively the momentum, position, frequency and mass of the $\alpha^{th}$ bath particle. The spin is coupled to the  bath via a coupling constant $\lambda_{\alpha j}$.\\
The time evolution of a quantum mechanical observable `$\boldsymbol{O}$' in the Heisenberg picture  can be written as \cite{cohen1986quantum}
\begin{equation}
    \Dot{O}=\frac{1}{i\hbar} [O,H]
    \label{heis}
\end{equation}
Our aim is to establish the quantum counterpart of the Langevin equations for the spin components \cite{Jayanavar1991,Seshadri1982} by making use of Eq.(\ref{heis}). We use the usual commutation and anticommutation relations for the spin components and also the canonical commutation relations for the bath variables in the derivation of the quantum Langevin equations \cite{sakurai2014modern}: 
\begin{eqnarray}
 \left[S_{i},S_{j}\right] &=&i\hbar \epsilon_{ijk}S_{k}
 \label{commutrel1}\\
 \lbrace S_{i},S_{j}\rbrace&=&\frac{\hbar^2}{2}\delta_{ij} I\label{anticommutrel1}\\
 \left[q_{i},p_{j}\right] &=& i\hbar \delta_{ij} I
 \label{commutrel2}
\end{eqnarray}
where $\epsilon_{ijk}$ is the Levi-Civita symbol, an anti-symmetric tensor connecting the spin components $S_{j}$s. The spin components are related to $M_j$s as $M_j=gS_j$, where $g$ is the gyromagnetic ratio given by $g=\frac{q}{m_q}$ ($q$ and $m_q$ are the charge and mass of the particle respectively.).\\

The equations of motion for the spin magnetic moments and the bath variables using two of the above  relations[Eq.(\ref{commutrel1}),(\ref{commutrel2})], can be written as\\ 
 \begin{align}
&\frac{dM_{z}}{dt} = g\sum_{\alpha}\left( \lambda_{\alpha x}q_{\alpha x}M_{y} -
  \lambda_{\alpha y}q_{\alpha y}M_{x}\right)  +\nonumber\\ &g\sum_{\alpha}\left(\frac{\lambda_{\alpha x}^{2}}{2m_{\alpha x}\omega_{\alpha x}^{2}} -\frac{\lambda_{\alpha y}^{2}}{2m_{\alpha y}\omega_{\alpha y}^{2}}\right)(M_x M_y+M_y M_x) \label{dMz}
 \end{align}
 \begin{align}
 &\frac{dM_{x}}{dt} = gH_{0}M_{y}+  g\sum_{\alpha}\left(\lambda_{\alpha y}q_{\alpha y}M_{z}-  \lambda_{\alpha z}q_{\alpha z}M_{y} \right)+\nonumber\\
 &g\sum_{\alpha} \left( \frac{\lambda_{\alpha y}^{2}}{2m_{\alpha y}\omega_{\alpha y}^{2}} -\frac{\lambda_{\alpha z}^{2}}{2m_{\alpha z}\omega_{\alpha z}^{2}}\right) \left( M_{y}M_{z}+M_{z}M_{y}\right) \label{dMx}
 \end{align}
 
 \begin{align}
 &\frac{dM_{y}}{dt} = -gH_{0}M_{x} +g\sum_{\alpha} \left(\lambda_{\alpha z}q_{\alpha z}M_{x} -\lambda_{\alpha x}q_{\alpha x}M_{z}\right)+\nonumber\\
 &g\sum_{\alpha}\left( \frac{\lambda_{\alpha z}^{2}}{2m_{\alpha z}\omega_{\alpha z}^{2}}  - \frac{\lambda_{\alpha x}^{2}}{2m_{\alpha x}\omega_{\alpha x}^{2}}\right) \left(M_{x}M_{z}+M_{z}M_{x}\right) \label{dMy}
 \end{align}
 \begin{eqnarray}
  \frac{dq_{\alpha j}}{dt} &=& \frac{p_{\alpha j}}{m_{\alpha j}}
  \label{qeq}
  \end{eqnarray}
  \begin{eqnarray}
  \frac{dp_{\alpha j}}{dt} &=& -m_{\alpha j} \omega_{\alpha j}^{2}q_{\alpha j}-\lambda_{\alpha j}M_{j} 
  \label{peq}
\end{eqnarray}
The equations for the bath variables are two coupled linear equations (Eq.(\ref{qeq}) and (\ref{peq})) and the solution for $q_{\alpha j}(t)$ is :
\begin{align}
    q_{\alpha j}(t) = q_{\alpha j}(0) cos(\omega_{\alpha j}t) + \frac{\Dot{q}_{\alpha j}(0)}{\omega_{\alpha j}} sin(\omega_{\alpha j}t) \nonumber \\
    -\frac{\lambda_{\alpha j}}{m_{\alpha j}\omega_{\alpha j}} \int_{0}^{t} sin[\omega_{\alpha j}(t-t^{'})] M_{j}(t^{'}) dt^{'} \label{bathsol}
\end{align}
On substituting the expression of $q_{\alpha j}(t)$ in the time evolution equation for the spin component $S_z$ (Eq.(\ref{dMz})) and using the anticommutation relation for the spin operators (Eq.(\ref{anticommutrel1})) we get
\begin{align}
\frac{dM_{z}}{dt} = gM_{y}\sum_{\alpha} \lambda_{\alpha x} (q_{\alpha x}(0) cos[\omega_{\alpha x}t]+
\frac{\Dot{q}_{\alpha x}(0)}{\omega_{\alpha x}}sin[\omega_{\alpha x}t]) -\nonumber\\
 gM_{y} \sum_{\alpha} \frac{\lambda_{\alpha x}^2}{m_{\alpha x}\omega_{\alpha x}} \int_{o}^{t}  sin[\omega_{\alpha x}(t-t^{'})] M_{x}(t^{'}) dt^{'} - \nonumber\\
gM_{x}\sum_{\alpha} \lambda_{\alpha y} (q_{\alpha y}(0) cos[\omega_{\alpha y}t]+
\frac{\Dot{q}_{\alpha y}(0)}{\omega_{\alpha y}}sin[\omega_{\alpha y}t]) +\nonumber\\
gM_{x} \sum_{\alpha} \frac{\lambda_{\alpha y}^2}{m_{\alpha y}\omega_{\alpha y}} \int_{o}^{t}  sin[\omega_{\alpha y}(t-t^{'})] M_{y}(t^{'}) dt^{'} 
\end{align}\\ \\
In a similar way, one can get the evolution equations for $M_{x}$ and $M_{y}$. \\
Evaluating the integral by parts in Eq.(11), we obtain 
\begin{align}
\frac{dM_{z}}{dt} =g M_{y}\sum_{\alpha} \lambda_{\alpha x} (q_{\alpha x}(0) cos[\omega_{\alpha x}t]+
\frac{\Dot{q}_{\alpha x}(0)}{\omega_{\alpha x}}sin[\omega_{\alpha x}t]) -\nonumber\\
g M_{y} \sum_{\alpha} \frac{\lambda_{\alpha x}^{2}}{m_{\alpha x}\omega_{\alpha x}^{2}} \int_{o}^{t}  cos[\omega_{\alpha x}(t-t^{'})] \Dot{M}_{x}(t^{'}) dt^{'} + \nonumber\\
 gM_{y} \sum_{\alpha}\frac{\lambda_{\alpha x}^{2}}{m_{\alpha x}\omega_{\alpha x}^{2}} cos[\omega_{\alpha x}t] M_{x}(0) -\nonumber\\
g M_{x}\sum_{\alpha} \lambda_{\alpha y} (q_{\alpha y}(0) cos[\omega_{\alpha y}t]+
\frac{\Dot{q}_{\alpha y}(0)}{\omega_{\alpha y}}sin[\omega_{\alpha y}t]) -\nonumber\\
g M_{x} \sum_{\alpha} \frac{\lambda_{\alpha y}^{2}}{m_{\alpha y}\omega_{\alpha y}^{2}} \int_{o}^{t}  cos[\omega_{\alpha y}(t-t^{'})] \Dot{M}_{y}(t^{'}) dt^{'} - \nonumber\\
gM_{x} \sum_{\alpha}\frac{\lambda_{\alpha y}^{2}
}{m_{\alpha y}\omega_{\alpha y}^{2}} cos[\omega_{\alpha y}t] M_{y}(0) + \nonumber\\
g\sum_{\alpha} \left(
\frac{\lambda_{\alpha x}^{2}}{m_{\alpha x}\omega_{\alpha x}^{2}}+\frac{\lambda_{\alpha y}^{2}}{m_{\alpha y}\omega_{\alpha y}^{2}}\right)M_{x}M_{y}
\end{align}
The time evolution equation for $M_{z}$ contains the initial values of $M_{x}$ and $M_{y}$. These values of $M_{x}(0)$ and $M_{y}(0)$ can be neglected 
in the long time limit \cite{Jayanavar1991}.\\
The quantum Langevin equations for all three spin components are written as 
\begin{align}
 \frac{dM_{z}}{dt} = gM_{y}(t) f_{x}(t) -gM_{x}(t) f_{y}(t)+\nonumber\\
 gM_{y}(t)\int_{0}^{t} \mu_{x}(t-t^{'})\Dot{M}_{x}(t^{'})dt^{'}- \nonumber\\ gM_{x}(t)\int_{0}^{t} \mu_{y}(t-t^{'})\Dot{M}_{y}(t^{'})dt^{'}+\nonumber\\
g \sum_{\alpha} \left(
\frac{\lambda_{\alpha x}^{2}}{m_{\alpha x}\omega_{\alpha x}^{2}}+\frac{\lambda_{\alpha y}^{2}}{m_{\alpha y}\omega_{\alpha y}^{2}}\right)M_{x}M_{y}
\label{eveqone}
\end{align}

\begin{align}
 \frac{dM_{x}}{dt} =gH_{0}M_{y}(t)- gM_{y}(t) f_{z}(t) +gM_{z}(t) f_{y}(t)-\nonumber\\
 g M_{y}(t)\int_{0}^{t} \mu_{z}(t-t^{'})\Dot{M}_{z}(t^{'})dt^{'} + \nonumber\\ g M_{z}(t)\int_{0}^{t} \mu_{y}(t-t^{'})\Dot{M}_{y}(t^{'})dt^{'}+\nonumber\\
 g \sum_{\alpha} \left(
\frac{\lambda_{\alpha y}^{2}}{m_{\alpha y}\omega_{\alpha y}^{2}}+\frac{\lambda_{\alpha z}^{2}}{m_{\alpha z}\omega_{\alpha z}^{2}}\right)M_{y}M_{z}
\label{eveqtwo}
\end{align}
\begin{align}
\frac{dM_{y}}{dt} =-g H_{0}M_{x}(t)- g M_{z}(t) f_{x}(t) +g M_{x}(t) f_{z}(t)+\nonumber\\
 g M_{x}(t)\int_{0}^{t} \mu_{z}(t-t^{'})\Dot{M}_{z}(t^{'})dt^{'} - \nonumber\\ g M_{z}(t)\int_{0}^{t} \mu_{x}(t-t^{'})\Dot{M}_{x}(t^{'})dt^{'}+\nonumber\\
 g \sum_{\alpha} \left(
\frac{\lambda_{\alpha z}^{2}}{m_{\alpha z}\omega_{\alpha z}^{2}}+ \frac{\lambda_{\alpha x}^{2}}{m_{\alpha x}\omega_{\alpha x}^{2}}\right)M_{z}M_{x}
\label{eveqthree}
\end{align}
where
\begin{align}
    f_{i} (t) = \sum_{\alpha} \lambda_{\alpha i} \left[ q_{\alpha x}(0) cos[\omega_{\alpha x}t]+
\frac{\Dot{q}_{\alpha x}(0)}{\omega_{\alpha x}}sin[\omega_{\alpha x}t] \right]
\end{align}

and
\begin{align}
    \mu_{i}(t-t^{'}) = \sum_{\alpha}\frac{\lambda_{\alpha i}^{2}}{m_{\alpha i}\omega_{\alpha i}^{2}} cos[\omega_{\alpha i}(t-t^{'})]
\end{align}
with $i=\{x,y,z\}$.

The quantum Langevin equations derived here for the spin components contain the random force functions $f_{i}$, which explicitly depend on 
the initial position and momentum coordinates of the bath particles.  The force functions have the following properties
\begin{equation}
    \langle f_{i}(t)\rangle = 0 
\end{equation}
\begin{align}
   \frac{1}{2}\langle\{f_{i}(t),f_{j}(t^{'})\}\rangle =\frac{\delta_{ij}}{2\pi} \int_{-\infty}^{+\infty} d\omega Re[K(\omega)] \hbar\omega\nonumber\\
   \times coth \left( \frac{\hbar \omega}{2k_{B}T}\right) e^{-i\omega (t-t^{'})}
\end{align}
with $i,j=\{x,y,z\}$ and $\delta_{ij}$ is the Kronecker delta function. $K (\omega) =\int_{-\infty}^{+\infty} dt K(t) e^{-i\omega t}$ and $K(t)=m \mu(t)$, where $m$ is the mass of the the spin-particle
and $\mu(t)$ is the memory kernel in the time domain. $k_B$ is the Boltzmann constant and $T$ is the temperature. We will make use of these properties of the force functions later on while studying the spin auto-correlations and the response function.\\

We have studied the spin dynamics incorporating two different types of heat-bath- the Ohmic bath and the Drude bath. In the next two sections, the spin auto-correlations are evaluated in detail for these two heat-baths.  We also 
evaluate the spin response function for the Ohmic bath.\\

In the classical limit, $M_i M_j$=$M_j M_i$ (the operators commute for $\hbar=0$) and we recover the results of  \cite{Jayanavar1991}:
\begin{align}
 \frac{dM_{z}}{dt} = gM_{y}(t) f_{x}(t) -gM_{x}(t) f_{y}(t)+ \nonumber \\
 gM_{y}(t)\int_{0}^{t} \mu_{x}(t-t^{'})\Dot{M}_{x}(t^{'})dt^{'}- \nonumber\\ gM_{x}(t)\int_{0}^{t} \mu_{y}(t-t^{'})\Dot{M}_{y}(t^{'})dt^{'}
\end{align}

\begin{align}
 \frac{dM_{x}}{dt} =gH_{0}M_{y}(t)- gM_{y}(t) f_{z}(t) +gM_{z}(t) f_{y}(t)-\nonumber\\
 g M_{y}(t)\int_{0}^{t} \mu_{z}(t-t^{'})\Dot{M}_{z}(t^{'})dt^{'} + \nonumber\\ g M_{z}(t)\int_{0}^{t} \mu_{y}(t-t^{'})\Dot{M}_{y}(t^{'})dt^{'}
\end{align}
\begin{align}
\frac{dM_{y}}{dt} =-g H_{0}M_{x}(t)- g M_{z}(t) f_{x}(t) +g M_{x}(t) f_{z}(t)+\nonumber\\
 g M_{x}(t)\int_{0}^{t} \mu_{z}(t-t^{'})\Dot{M}_{z}(t^{'})dt^{'} - \nonumber\\ g M_{z}(t)\int_{0}^{t} \mu_{x}(t-t^{'})\Dot{M}_{x}(t^{'})dt^{'}
\end{align}
\subsection{Ohmic bath}
\vspace{0.2cm}
The above equations for the spin components (Eq.\ref{eveqone}-\ref{eveqthree}) are all non-Markovian. However, these equations can be solved in the Markovian limit by using the Ohmic bath model. The Ohmic bath is defined in terms of a delta correlated memory kernel :  
\begin{equation}
    \mu_{i}(t-t^{'})= 2\gamma \delta(t-t^{'})
    \label{ohmicbath}
\end{equation}
with $\gamma$ the damping coefficient.\\
One can equivalently define the Ohmic bath in terms of the spectral density of the coupling
to the bath \cite{Caldeira}: 
$J(\omega)=\gamma \omega$.
We introduce an upper cutoff frequency $\Omega$ above which $J(\omega)$ vanishes \cite{Caldeira}. 
Using this, the above set of equations  (Eq.\ref{eveqone}-\ref{eveqthree})  take the form 
\begin{align}
   &\frac{dM_{z}}{dt} =g M_{y}(t) f_{x}(t) -g M_{x}(t) f_{y}(t)+
2\gamma g  M_{y}(t) \Dot{M}_{x}(t)+ \nonumber\\ 
&2\gamma g M_{x}(t)\Dot{M}_{y}(t)+
  2\Omega\gamma g M_{x}(t)M_{y}(t)\\
 &\frac{dM_{x}}{dt} = g H_{0}M_{y}(t)- g M_{y}(t)f_{z}(t) +g M_{z}(t) f_{y}(t)-  \nonumber\\
&2\gamma g M_{y}(t) \Dot{M}_{z}(t)+ 2\gamma g M_{z}(t)\Dot{M}_{y}(t)+
  2 \Omega \gamma g M_{y}(t)M_{z}(t)\\
&\frac{dM_{y}}{dt} = -g H_{0}M_{x}(t)-g M_{z}(t)f_{x}(t) +g M_{x}(t) f_{z}(t)+ \nonumber\\
&2\gamma g  M_{x}(t) \Dot{M}_{z}(t)- 2\gamma g M_{z}(t)\Dot{M}_{x}(t)+
  2 \Omega\gamma g M_{z}(t)M_{x}(t)
\end{align}
where
\begin{align}
\frac{\lambda_{\alpha x}^{2}}{m_{\alpha x}\omega_{\alpha x}^{2}} = \frac{\lambda_{\alpha y}^{2}}{m_{\alpha y}\omega_{\alpha y}^{2}}= \Omega \gamma   
\end{align}
We now evaluate the spin auto-correlation function. We first consider the expectation value of 
the $z-$component of the spin magnetic moment to be a constant  i.e. $\langle M_{z}\ra = M_{z}=$ constant \cite{Dudarev2011}. \\

At equilibrium, the z-component of the spin magnetic moment along the direction of the magnetic field can be calculated as the statistical average of the spin magnetic moment along the magnetic field via the Gibbs distribution. It is given by \cite{kittel1996introduction}:
\begin{align}
M_z=\sum_{m_s=-S}^{m_s=+S}\frac{g m_s e^{g m_s H_0/k_BT}}{e^{g m_s H_0/k_BT}} \label{tempdep}
\end{align}
After some algebraic manipulations, Eq.(\ref{tempdep}) gives,
\begin{align}
    M_z=-gSB_s(x) \label{Gibbsdist}
\end{align}
where, $x=\frac{SH_{0}}{k_{B}T}$ and $B_{s}(x)$ is called the Brillouin function given by \cite{kittel1996introduction,darby1967tables}:
\begin{align}
    B_{s}(x)=\frac{2S+1}{2S}coth \left[ \left(\frac{2S+1}{2S} \right) x \right]- \frac{1}{2S}coth \left[ \left(\frac{1}{2S} \right) x \right] \label{Brillouinfunc}
\end{align}
$B_{s}(x)$ takes different forms in the high and low temperature limits (See the Appendix for details).
Within this approximation, the time evolution equations of the expectation values of the spin components take the following 
form : 
\begin{align}
 \frac{d\langle M_{x}\rangle}{dt} = g H_{0}\langle M_{y}(t)\rangle+ 2\gamma g M_{z}(t)\frac{d\langle M_{y}(t)\rangle}{dt}\nonumber\\
+2\Omega\gamma g M_{z}(t)\langle M_{y}(t)\rangle \label{coup1}
\end{align}
\begin{align}
  \frac{d\langle M_{y}\rangle}{dt} = -g H_{0}\langle M_{x}(t)\rangle- 2\gamma g M_{z}(t)\frac{d\langle M_{x}(t)\rangle}{dt}\nonumber\\
  +2 \Omega\gamma g M_{z}(t)\langle M_{x}(t)\rangle\label{coup2}
\end{align}
Solving the two coupled equations (Eqs.(\ref{coup1}) and (\ref{coup2})), we arrive at
\begin{align}
   \langle M_{x}\rangle =&M_{xy}  exp\left(-\frac{2M_{z}\gamma g^2 t (H_{0}+2M_{z}
   \Omega \gamma)}{1+4M_{z}^{2}g^{2} \gamma^{2}}\right)\times \nonumber \\ &cos\left(\frac{gt(H_{0}+2M_{z}\Omega \gamma)}{1+4M_{z}^{2}\gamma^{2} g^{2}}-\phi\right)
  \end{align}
  \begin{align}
    \langle M_{y}\rangle = &M_{xy} exp \left(-\frac{2M_{z}\gamma g^{2} t (H_{0}+2M_{z} \Omega \gamma)}{1+4M_{z}^{2} g^{2}\gamma^{2}}\right) \times \nonumber \\ &sin\left(\frac{gt(H_{0}+2M_{z}\Omega \gamma)}{1+4M_{z}^{2}\gamma^{2} g^{2}}-\phi\right)
\end{align}
with $M_{xy}=\sqrt{M^{2}-M_{z}^{2}}$, $cos \phi = \frac{\langle M_{x}(0)\rangle}{\sqrt{M^{2}-M_{z}^{2}}}$ and $sin \phi = \frac{\langle M_{y}(0)\rangle}{\sqrt{M^{2}-M_{z}^{2}}}$, where M is the total spin magnetic moment.\\

\subsubsection{Spin auto-correlation function}
\vspace{0.1cm}
In order to calculate the spin auto-correlation function, we use the following approximation which is valid for a spin precessing rapidly about the $z$ direction, which is the direction of the applied magnetic field \cite{Dudarev2011} :  $\langle M_{i}(t)M_{i}(0)\rangle\approx\langle M_{i}(t) \rangle \langle M_{i}(0)\rangle$ with $i={x,y}$ and $\langle M_{z}(t)M_{z}(0)\rangle= {M}_{z}^{2}$. Thus the spin auto-correlation function is 
\begin{equation}
   C(t)= \langle M(t).M(0)\rangle = \langle M_{x}(t)M_{x}(0)\rangle +\langle M_{y}(t)M_{y}(0)\rangle+ {M}_{z}^{2}
   \label{spcorr1}
\end{equation}

The explicit analytical form of the spin auto-correlation function is :
\begin{equation}
\begin{split}
   C(t)=& \langle M(t).M(0)\rangle \\=&
   {M}_{z}^{2} +(M^{2}-M_{z}^{2}) e^{-t/\tau_{R}} cos\left[\frac{g (H_0+2M_{z}\Omega \gamma)}{1+4M_{z}^{2}g^{2}\gamma^{2}}t\right]\\
   =&
   {M}_{z}^{2} +(M^{2}-M_{z}^{2}) e^{-t/\tau_{R}} cos\left[\tilde{\omega}_Lt\right]
   \label{spincorr}
   \end{split}
\end{equation} 
with the relaxation time of the spin auto-correlation function given by $\tau_{R}=\frac{1+4M_{z}^{2}g^2 \gamma^{2}}{2M_{z}\gamma g^2(H_{0}+2M_{z}\Omega \gamma)} $.\\
$gH_0$ is the Larmour frequency ($\omega_L$) and $\tilde{\omega}_L$ is given by:
\begin{align}
  \tilde{\omega}_L= \frac{g (H_0+2M_{z}\Omega \gamma)}{1+4M_{z}^{2}g^{2}\gamma^{2}}  
\end{align}
Notice that the relaxation time $\tau_R$ corresponds to the relaxation of the $x$ and $y$ components of the spin due to dissipation in the presence of the bath particles.\\ 
Since the magnetic moment $M_z$ discussed in the last section is temperature dependent, 
the relaxation time $\tau_{R}$, which is a function of $M_z$, also varies with
temperature.\\

In the classical limit and for small $\gamma$, one can get \cite{Dudarev2011},
\begin{align}
   C(t) =M_{z}^{2}+(M^{2}-M_{z}^{2})  e^{-t/\tau_{R(Cl)}} cos\left[gH_{0} t\right] \label{corrclassical}
  \end{align}
  where $\tau_{R(Cl)}=\frac{1}{2 M_z \gamma g^2 H_0}$. \\ \\
Notice that the relaxation time ($\tau_{R}$) goes to zero for $\Omega$ $\rightarrow \infty$, unless $M_{z}$ or $\gamma$ is extremely small. Thus we can see that at very large values of $\Omega$, the spin magnetic moments and the correlation functions decay very fast for large $\gamma$ and higher values of $M_{z}$ i.e. at low temperatures (see Eq.(\ref{Gibbsdist})). This point is further discussed in the following section.
\subsubsection{Response function}
\vspace{0.1cm}
The spin response function $R(t)$ is defined via the following relation: 
\begin{equation}
    <M(t)> = \int_{-\infty}^{t}{R(t-t')f(t')dt'}
    \label{resp}
\end{equation}
where $M(t)$ is the magnetisation and $f(t)$ is the perturbing magnetic field. 
The quantum Fluctuation-Dissipation theorem in the frequency domain states that \cite{kubo1966fluctuation}
\begin{equation}
    \frac{1}{\hbar}C(\omega) = coth\left(\frac{\omega}{\Omega_{th}}\right) R^{''}(\omega)
    \label{fdt}
\end{equation}
where $\Omega_{th}=\frac{2k_B T}{\hbar}$ . $R^{''}(\omega)$ is the imaginary part of the total spin response function.\\
Thus $R^{''}(\omega)$ can be calculated from  Eq.(\ref{spincorr}) and Eq.(\ref{fdt}). We obtain 
the imaginary part $R^{''}(t)$ of the response function  in the time domain via an inverse Fourier transform : 
\begin{align}
   &R^{''}(t)=\nonumber\\
    &\frac{M_{xy}^2 e^{-(A+iB)t}\left(tanh(\frac{2\left(B-iA\right)}{\Omega_{th}})-e^{2iBt}tanh(\frac{2\left(B+iA \right)}{\Omega_{th}})\right)}{2\hbar} \label{ImR}
\end{align}
where,
\begin{align}
   & A=\frac{2M_z \gamma g^2\left(H_0+2M_z \Omega \gamma \right)}{1+4 M_z^2 g^2 \gamma^2}\\
   &B= \frac{g \left(H_0+2M_z \Omega \gamma \right)}{1+4 M_z^2 g^2\gamma^2}
\end{align}

Now, one can derive the real part of the Response Function in the Fourier domain ($R^{'}(\omega)$) from $R^{''}(\omega)$, using the Kramers-Kronig transformation given by \cite{de1926theory,kramers1927diffusion}:
\begin{align}
R^{'}(\omega)=\frac{1}{\pi}\int_{-\infty}^{\infty}\frac{\omega^{'} R^{''}(\omega^{'})}{(\omega^{'2}-\omega^2)}d\omega^{'}\label{Kramerskronig}
\end{align}
Here we make use of Eq.(\ref{Kramerskronig}) and obtain $R^{'}(\omega)$:
\begin{equation}
\begin{split}
    R^{'}(\omega)=&\frac{M_{xy}^2}{\sqrt{2\pi}}\bigg(\frac{tanh\left(\frac{2(B-iA)}{ \Omega_{th}} \right) \left(B-iA\right)}{\hbar\left[(B-iA)^{2}-\omega^{2}\right]} +\\
   & \frac{tanh\left(\frac{2(B+iA)}{ \Omega_{th}} \right) \left(B+iA\right)}{\hbar\left[(B-\omega+iA)(B+\omega+iA)\right]} \bigg) 
    \end{split}
\end{equation}
Hence $R^{'}(t)$ is:
\begin{align}
&R^{'}(t)=-iM_{xy}^{2} e^{-(A+iB)t}\times\nonumber\\
&\frac{
   \left(tanh(\frac{2\left(iA-B\right)}{\Omega_{th}})+e^{2iBt}tanh(\frac{2\left(B+iA \right)}{\Omega_{th}})\right)}{2\hbar} \label{ReR}
\end{align}

Therefore, the total spin response function in the time domain is:
\begin{equation}
    R(t)=R^{'}(t)+iR^{''}(t) \label{totresp}
\end{equation}
The explicit expression for the total spin response function in the time domain
obtained using Eq.(\ref{ImR}), Eq.(\ref{ReR})and Eq.(\ref{totresp}) is given by:
\begin{align}
&R(t)= -2M_{xy}^{2}e^{-A t} \times \nonumber\\
    &\frac{\left(cos(Bt)sin\left(\frac{4A}{\Omega_{th}}\right) +sin(Bt)sinh\left(\frac{4B}{\Omega_{th}}\right) \right)}{\hbar \left( cos \left(\frac{4A}{\Omega_{th}}  \right) +cosh \left(\frac{4B}{\Omega_{th}}  \right) \right)}
\end{align}
\subsection{Drude bath}
\vspace{0.2cm}
In the last section, we have used a delta correlated memory kernel pertaining to an Ohmic bath and have evaluated the expectation values of the spin components, spin auto-correlation and the spin response function.  We can go beyond 
the Ohmic model and derive time evolution equations for the spin components which can be solved in the non-Markovian limit using a kernel with a non-trivial memory, for instance the Drude bath model. The memory kernel for the Drude bath is 
\begin{equation}
    \mu(t-t^{'})= \frac{\gamma}{\tau}e^{-\left(\frac{t-t^{'}}{\tau}\right)} \Theta(t)
\end{equation}
with $\tau$ the memory time and $\Theta(t)$ the Heaviside Theta Function.\\
Using the above kernel and rearranging the equations for the $x$ and $y$-components of the spin(Eq.(15)-(16)) we obtain the time evolution equations: 
\begin{align}
&\frac{dM_{x}}{dt} =g H_{0}M_{y}(t)-g M_{y}(t)f_{z}(t)+g M_{z}(t)f_{y}(t)\nonumber\\
&-\frac{g \gamma}{\tau}M_{y}(t)\int_{0}^{t} e^{-\frac{(t-t^{'})}{\tau}} \Dot{M}_{z}(t^{'})dt^{'}\nonumber\\
&+\frac{g \gamma}{\tau}M_{z}(t)\int_{0}^{t}e^{-\frac{(t-t^{'})}{\tau}} \Dot{M}_{y}(t^{'})dt^{'} 
+\frac{2g \gamma}{\tau}M_{y}(t)M_{z}(t) \label{Drude1}
\end{align}
\begin{align}
&\frac{dM_{y}}{dt} = -g H_{0}M_{x}(t)-g M_{z}(t)f_{x}(t)+g M_{x}(t)f_{z}(t)+\nonumber\\
& \frac{g \gamma}{\tau}M_{x}(t)\int_{0}^{t} e^{-\frac{(t-t^{'})}{\tau}} \Dot{M}_{z}(t^{'})dt^{'}\nonumber\\
&-\frac{g \gamma}{\tau}M_{z}(t)\int_{0}^{t} e^{-\frac{(t-t^{'})}{\tau}} \Dot{M}_{x}(t^{'})dt^{'} +\frac{2g \gamma}{\tau}M_{z}(t)M_{x}(t) \label{Drude2}
\end{align}
Now, multiplying both sides of Eqs. (\ref{Drude1}) and (\ref{Drude2}) by $e^{t/\tau}$, one gets,
\begin{align}
    &e^{t/\tau}\frac{dM_{x}}{dt} =e^{t/\tau}\Big[g H_{0}M_{y}(t)-g M_{y}(t)f_{z}(t)+g M_{z}(t)f_{y}(t)  \nonumber\\
    &+\frac{2g \gamma}{\tau} M_{y}(t)M_{z}(t)\Big]
-\frac{g \gamma}{\tau}M_{y}(t)\int_{0}^{t} e^{\frac{t'}{\tau}} \Dot{M}_{z}(t^{'})dt^{'}\nonumber\\
&+\frac{g \gamma}{\tau}M_{z}(t)\int_{0}^{t}e^{\frac{t'}{\tau}} \Dot{M}_{y}(t^{'})dt^{'} 
\label{Drude3}
\end{align}
\begin{align}
    &e^{t/\tau}\frac{dM_{y}}{dt} =e^{t/\tau}\Big[-g H_{0}M_{x}(t)-g M_{z}(t)f_{x}(t)+g M_{x}(t)f_{z}(t)  \nonumber\\
    &+\frac{2g \gamma}{\tau} M_{z}(t)M_{x}(t)\Big]
+\frac{g \gamma}{\tau}M_{x}(t)\int_{0}^{t} e^{\frac{t'}{\tau}} \Dot{M}_{z}(t^{'})dt^{'}\nonumber\\
&-\frac{g \gamma}{\tau}M_{z}(t)\int_{0}^{t}e^{\frac{t'}{\tau}} \Dot{M}_{x}(t^{'})dt^{'}\label{Drude4} 
\end{align}

Taking a time derivative on both sides of the above two equations (Eqs.(\ref{Drude3}) and (\ref{Drude4})) and then taking an expectation value we arrive at the following 
equations, keeping in mind that the expectation value of the $z-$component of the spin is a constant \cite{Dudarev2011}.
\begin{align}
   & \frac{d^{2}\langle M_{x}\rangle}{dt^{2}} +\frac{1}{\tau} \frac{d\langle M_{x}\rangle}{dt}-g \left(H_{0}+\frac{3\gamma}{\tau}M_z\right) \frac{d\langle M_{y}\rangle}{dt}\nonumber\\
   & - \frac{g}{\tau}\left(H_{0}+\frac{2\gamma}{\tau} M_z\right) \langle M_{y}\rangle=0
   \end{align}
\begin{align}
& \frac{d^{2}\langle M_{y}\rangle}{dt^{2}} +\frac{1}{\tau} \frac{d\langle M_{y}\rangle}{dt}+g \left(H_{0}+\frac{3\gamma}{\tau}M_z\right) \frac{d\langle M_{x}\rangle}{dt}\nonumber\\
& + \frac{g}{\tau}\left(H_{0}+\frac{2\gamma}{\tau} M_z\right) \langle M_{x}\rangle=0 
\end{align}
On solving the above two equations, one can obtain the expectation values values of $M_{x}$ and $M_{y}$ as follows:
\begin{align}
&\langle M_{x} \rangle = \frac{1}{2} \langle M_{x}(0) \rangle \Bigg[\frac{e^{-\frac{(a+ib)t}{2}} }{\sqrt{(a+ib)^{2}-4ic}} \times\nonumber\\
&cosh\left(\frac{\sqrt{(a+ib)^{2}-4ic}}{2} t+\phi_{1} \right) + \frac{e^{-\frac{(a-ib)t}{2}} }{\sqrt{(a-ib)^{2}+4ic}} \times \nonumber\\
&cosh\left(\frac{\sqrt{(a-ib)^{2}+4ic}}{2} t+\phi_{2} \right)\bigg] \\
&\langle M_{y} \rangle = \frac{1}{2} \langle M_{y}(0) \rangle \Bigg[\frac{e^{-\frac{(a+ib)t}{2}} }{\sqrt{(a+ib)^{2}-4ic}} \times\nonumber\\
&cosh\left(\frac{\sqrt{(a+ib)^{2}-4ic}}{2} t+\phi_{1} \right) + \frac{e^{-\frac{(a-ib)t}{2}} }{\sqrt{(a-ib)^{2}+4ic}} \times \nonumber\\
&cosh\left(\frac{\sqrt{(a-ib)^{2}+4ic}}{2} t+\phi_{2} \right)\bigg] 
\end{align} \\ 
with $a=\frac{1}{\tau}$, $b=g(H_{0}+\frac{3\gamma}{\tau}M_z)$, $c=\frac{g}{\tau}(H_{0}+\frac{2\gamma}{\tau}M_z)$ and $\phi_{1}=tan^{-1}\big[\frac{a+ib}{\sqrt{(a+ib)^{2}-4ic)}}\big]$; $\phi_{2}=tan^{-1}\Big[\frac{a-ib}{\sqrt{(a-ib)^{2}+4ic)}}\Big]$. 
\subsubsection{Spin auto-correlation function}
The spin auto-correlation for the Drude bath model is now evaluated following a method similar to the one 
used for the Ohmic bath(see Eq.(\ref{spcorr1})): 
\begin{align}
    &C(t)= M_z^{2} +  \frac{1}{2}\left( \langle M_{x}(0) \rangle ^{2} +\langle M_{y}(0) \rangle^{2} \right)\times \nonumber\\
    & \Bigg[\frac{e^{-\frac{(a+ib)t}{2}} }{\sqrt{(a+ib)^{2}-4ic}}cosh\left(\frac{\sqrt{(a+ib)^{2}-4ic}}{2} t+\phi_{1} \right) \nonumber\\ 
    &+ \frac{e^{-\frac{(a-ib)t}{2}} }{\sqrt{(a-ib)^{2}+4ic}} 
cosh\left(\frac{\sqrt{(a-ib)^{2}+4ic}}{2} t+\phi_{2} \right)\bigg]
\end{align}
All the parameters (e.g. $a,b,c,\phi_{1},\phi_{2}$) are defined in the previous sub-section.\\
As in the Ohmic case, here also we notice a damped oscillatory 
behaviour of the various functions, with the damping rate being determined by the factors $H_0$, T, $\gamma$ and $\tau$. The detailed results and the characteristic behaviours corresponding to different values of memory time $\tau$ are given in the following section.
\section{Results and Discussion}
\vspace{0.3cm}
In Figs.(\ref{fig1}-\ref{fig3}), we have plotted the $x$ component ($\langle M_x(t)\rangle$) and the $y$ component ($\langle M_y(t)\rangle$) of the expectation values of magnetic moments for a particle with spin angular momentum $ \hbar/2$, the spin auto-correlation function ($C(t)$) and the response function ($R(t)$) as a function of time for both the classical high temperature and the quantum low temperature domains corresponding to the under-damped ($\tilde{\omega}_L>1/\tau_{R}$) and the over-damped ($\tilde{\omega}_L<1/\tau_{R}$) regimes for the Ohmic bath model.  Figs.(\ref{fig4}-\ref{fig5}) represent the plots of $\langle M_x(t)\rangle$ and $\langle M_y(t)\rangle$, $C(t)$ for the Drude bath model with a finite memory time. The evaluation of the response function for the Drude bath model involves complicated 
functions which makes the calculation cumbersome. We have therefore restricted to the Ohmic
bath model for the evaluation of the response function. \\
Fig.(\ref{fig1}[a]) shows that $\langle M_x(t)\rangle$ and $\langle M_y(t)\rangle$ exhibit damped oscillations in the under damped regime. Fig(\ref{fig1}[b]) represents the over-damped regime, where one notices a monotonic decay of $\langle M_x(t)\rangle$ and $\langle M_y(t)\rangle$ due to the high value of the damping coefficient. In the low temperature quantum domain, the oscillations for the under-damped regime are more sustained followed by a slower decay of $\langle M_x(t)\rangle$ and $\langle M_y(t)\rangle$ (Figs.\ref{fig2}[c-d]), whereas, we notice that the magnetic moments at low temperatures and high damping fall off to extremely low values over a very short time for the Ohmic model, as described in the previous section (see Fig.\ref{fig1}[d]). This is due to the fact that the relaxation time $\tau_{R} \rightarrow$0 when $\gamma$ is large and temperature is low ($M_z$ is high).    \\
Similar plots are obtained for the spin auto-correlation functions, for the high and the low temperature domains as shown in Figs.(\ref{fig2}[c]) and (\ref{fig2}[d]). The spin auto-correlation function decays with increasing time and settles to a constant value given by the z-component of the spin auto-correlation function, determined by the Gibbs distribution at thermal equilibrium.\\
We have also plotted the response function for the Ohmic model in Fig.(\ref{fig3}) for the high temperature classical and the low temperature quantum domains. The response function increases with time, when the perturbing field is turned on and then reaches a maximum and again gets damped as a result of coupling with the bath particles. Damped oscillations are present in the under-damped cases, as seen in Figs.(\ref{fig3}[a]) and (\ref{fig3}[c]). However, one notices that these oscillations are more pronounced in the low temperature case (see Fig.\ref{fig3}[c]) compared to the high temperature case (see Fig.(\ref{fig3}[a]).\\
Figs.(\ref{fig4}) and (\ref{fig5}) represent the expectation values of the spin components and the spin auto-correlation functions respectively for the Drude bath model. Figs.(\ref{fig4}[a]) and (\ref{fig4}[c]) show that the oscillations in $\langle M_x(t)\rangle$ and $\langle M_y(t)\rangle$ die off faster and they decay quickly relative to the low temperature cases (see Figs.(\ref{fig4}[c]) and (\ref{fig4}[d])), which is expected  due to the fact that spin alignments are lost faster at higher temperatures. \\
Moreover, here we have presented the results for two different memory times ($\tau=0.1$ and $\tau=5$) and it can be clearly seen that the decay rates of the spin components and also the spin auto-correlation functions are visibly lower for higher values of the memory time in the Drude bath model.\\
In Figs.(\ref{contour_plot})[a-c], we have shown the regions of different relaxation time $\tau_{R}$ for the Ohmic model as a function of: \textbf{[a]} the  temperature ($T$) and the magnetic field ($H_{0}$); \textbf{[b]} the temperature ($T$) and the damping coefficient ($\gamma$); and \textbf{[c]} the magnetic field ($H_{0}$) and the damping coefficient ($\gamma$). We have already mentioned that the magnetic moments and the correlation functions decay fast for a large 
value of $M_z$ or $\gamma$. This is simply due to the fact that $\tau_{R} \rightarrow 0$ for large  $M_z$ and $\gamma$, which is evident from Figs. (\ref{contour_plot})[a-c]. It is to be noted that the spin tends to align along the direction of the field leading to an increase in the magnetic moment in the direction of $H_0$ ($i.e.$ $M_{z}$),
with an increase in the magnetic field and a reduction in the temperature (see Eq.(\ref{Gibbsdist}).  The plots in (Fig.(\ref{contour_plot})) highlight this trend of the relaxation time of the oscillating spin in terms of the varying magnetic field, the temperature and the damping coefficient. \\
In the plots we have considered $M_x(0)=\sqrt{\frac{3}{5}}M_{xy}$, $M_y(0)=\sqrt{\frac{2}{5}}M_{xy}$, $\hbar$=1 and $k_B$=1. The magnetic moments, the correlation functions and the response functions are in units of the Bohr magneton ($\mu_B$), $\mu_B^2$ and $\mu_B^2/\hbar$  respectively, whereas, $T$, $H_0$, $\gamma$ and the frequencies are expressed in general S.I. units.

\begin{widetext}

\begin{figure}[H]
     \centering
     
     \includegraphics[scale=0.445]{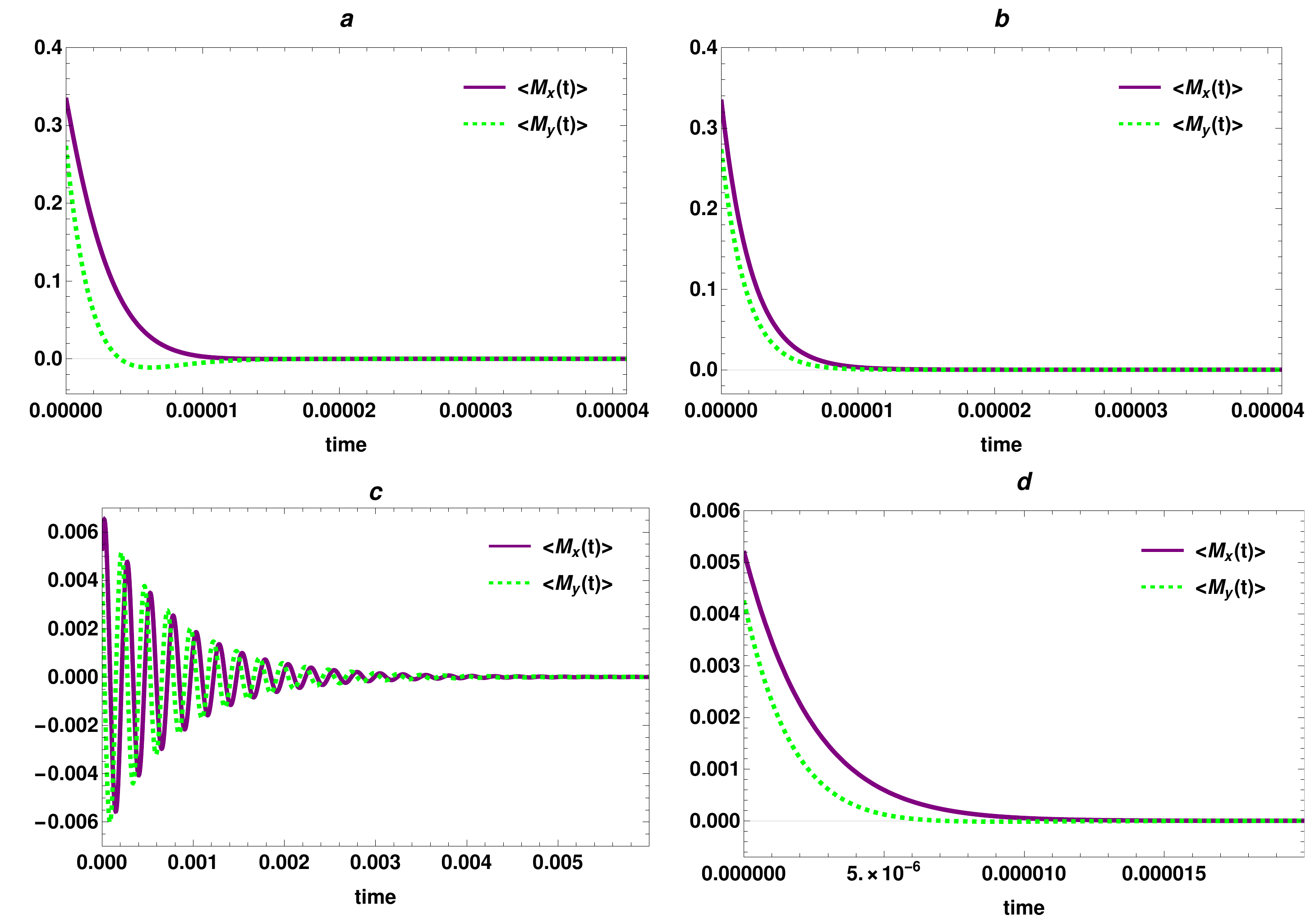}
     
     \caption{The $x$ and $y$ components of the expectation values of the magnetic moments ($\mu_B$) for the Ohmic model are plotted as a function of time for the under-damped ($\tilde{\omega}_L>1/\tau_{R}$) and the over-damped ($\tilde{\omega}_L<1/\tau_{R}$) regimes corresponding to the classical high temperature domain and the quantum low temperature domain with S=1/2, g=1 and $\Omega=10^6$: \textbf{[a]} T=10, $H_0 = 8$, $\gamma = 5$, \textbf{[b]} T=10, $H_0=8$, $\gamma=20 $, \textbf{[c]} T=0.01, $H_0=0.1 $, $\gamma=0.05 $  and  \textbf{[d]} T=0.01, $H_0=0.1 $, $\gamma=5 $. }
     \label{fig1}
 \end{figure}

\begin{figure}[H]
     \centering
     \includegraphics[scale=0.449]{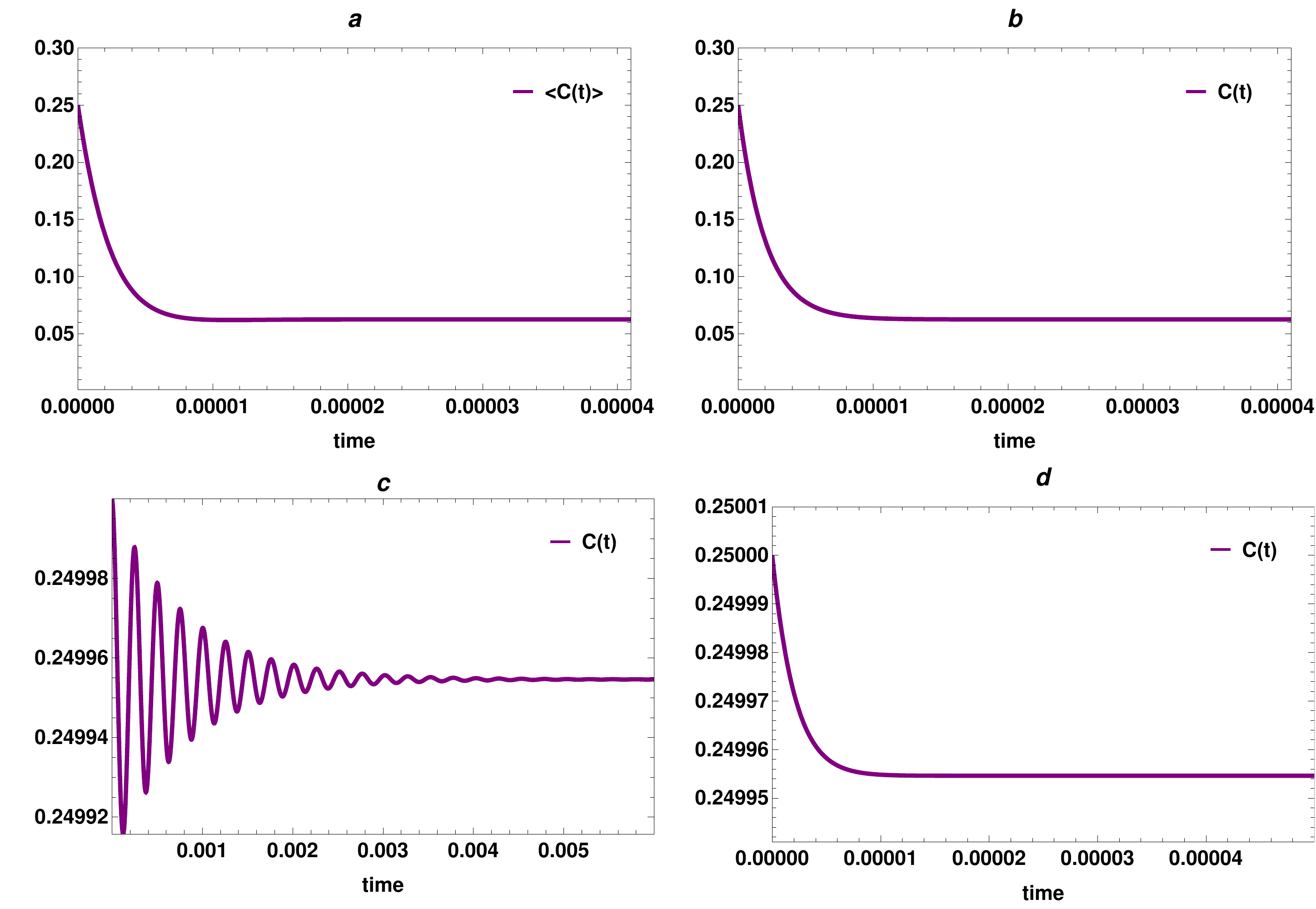}
    \caption{The spin auto-correlation function ($\mu_B^2$) for the Ohmic model is plotted as a function of time for the under-damped ($\tilde{\omega}_L>1/\tau_{R}$) and the over-damped ($\tilde{\omega}_L<1/\tau_{R}$) regimes corresponding to the classical high temperature domain and the quantum low temperature domain with S=1/2, g=1 and $\Omega=10^6$: \textbf{[a]} T=10, $H_0 = 8 $, $\gamma = 5 $, \textbf{[b]} T=10, $H_0=8 $, $\gamma=20 $, \textbf{[c]} T=0.01, $H_0=0.1 $, $\gamma=0.05 $  and  \textbf{[d]} T=0.01, $H_0=0.1 $, $\gamma=5 $.}
     \label{fig2}
 \end{figure}

\begin{figure}[H]
     \centering
     \includegraphics[scale=0.465]{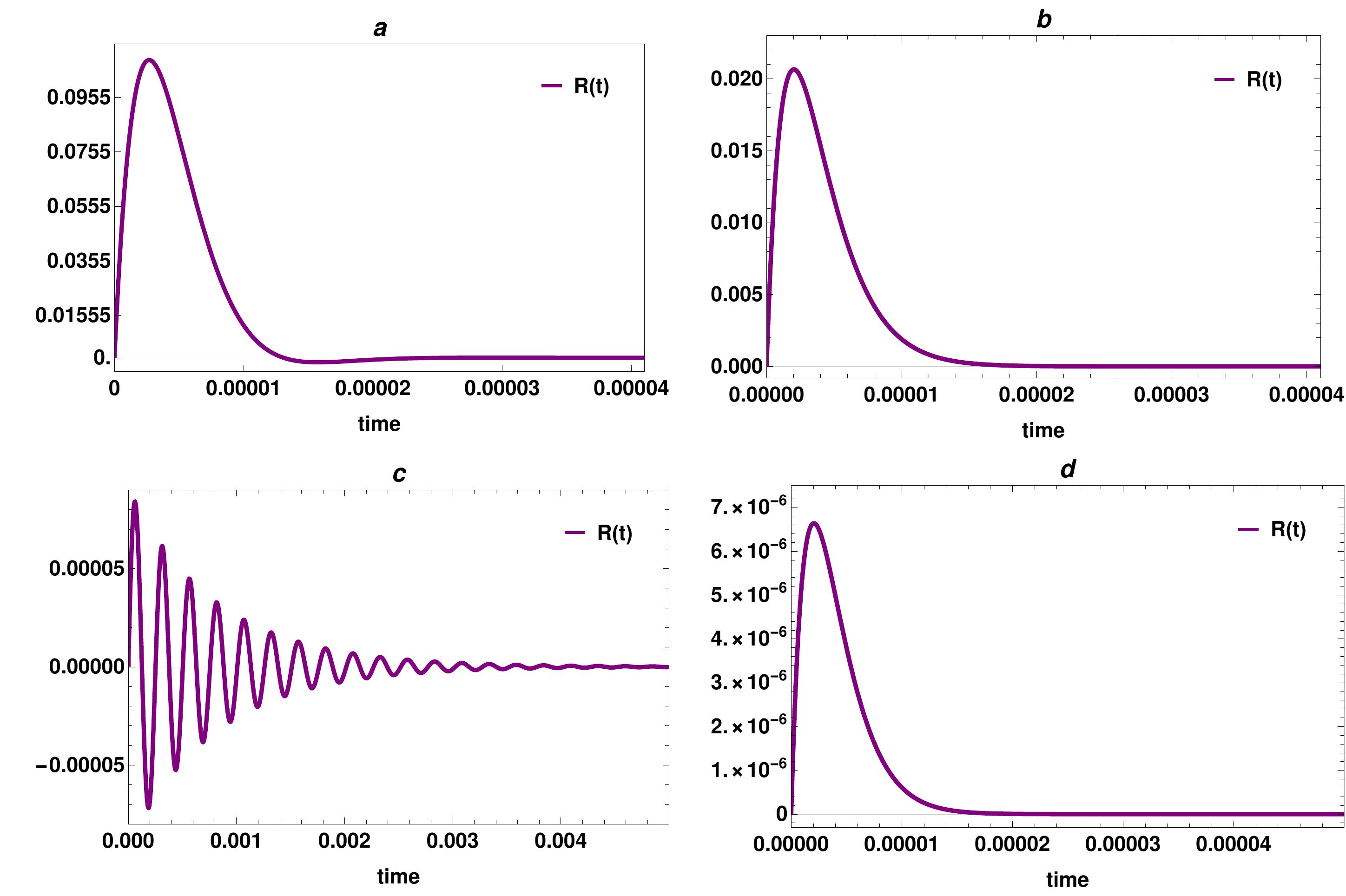}
     \caption{The response function ($\mu_B^2/\hbar$) for the Ohmic model is plotted as a function of time for the under-damped ($\tilde{\omega}_L>1/\tau_{R}$) and the over-damped ($\tilde{\omega}_L<1/\tau_{R}$) regimes corresponding to the classical high temperature domain and the quantum low temperature domain with S=1/2, g=1 and $\Omega=10^6$: \textbf{[a]} T=10, $H_0 = 8 $, $\gamma = 5 $, \textbf{[b]} T=10, $H_0=8 $, $\gamma=20 $, \textbf{[c]} T=0.01, $H_0=0.1 $, $\gamma=0.05 $  and  \textbf{[d]} T=0.01, $H_0=0.1 $, $\gamma=5 $.}
     \label{fig3}
 \end{figure}

\begin{figure}[H]
     \centering
     \hspace{0.2cm}
     \includegraphics[scale=0.44]{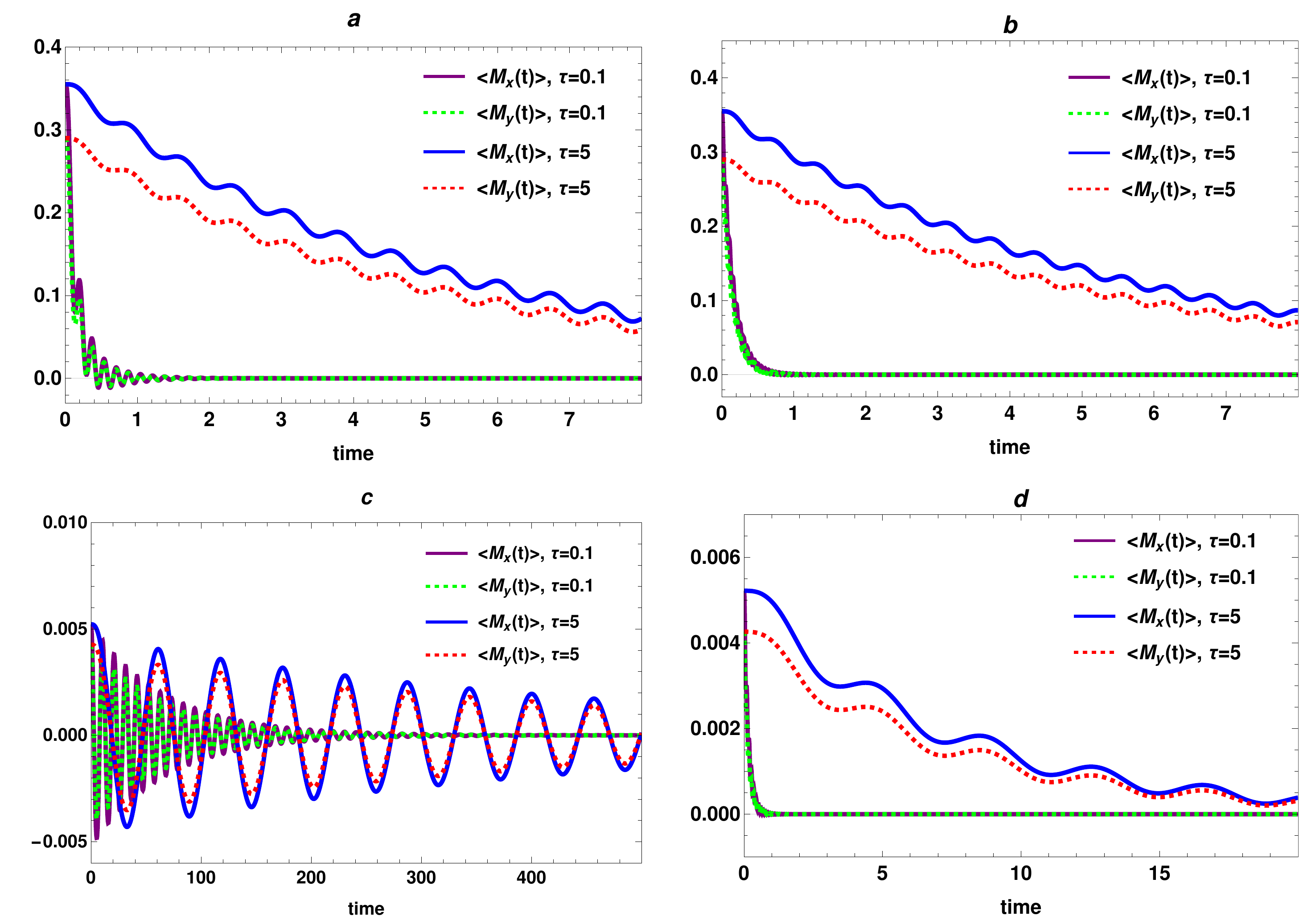}
    \caption{The x and y components of the expectation values of the magnetic moments ($\mu_B$) for the Drude model (memory times $\tau$=0.1 and 5) are plotted as a function of time for the under-damped (low $\gamma$) and the over-damped (high $\gamma$) regimes corresponding to the classical high temperature domain and the quantum low temperature domain with S=1/2, g=1 and $\Omega=10^6$: \textbf{[a]} T=10, $H_0 = 8 $, $\gamma = 5 $, \textbf{[b]} T=10, $H_0=8 $, $\gamma=20 $, \textbf{[c]} T=0.01, $H_0=0.1 $, $\gamma=0.05 $  and  \textbf{[d]} T=0.01, $H_0=0.1 $, $\gamma=5 $.}
     \label{fig4}
 \end{figure}
 
  \begin{figure}[H]
     \centering
     \includegraphics[scale=0.45]{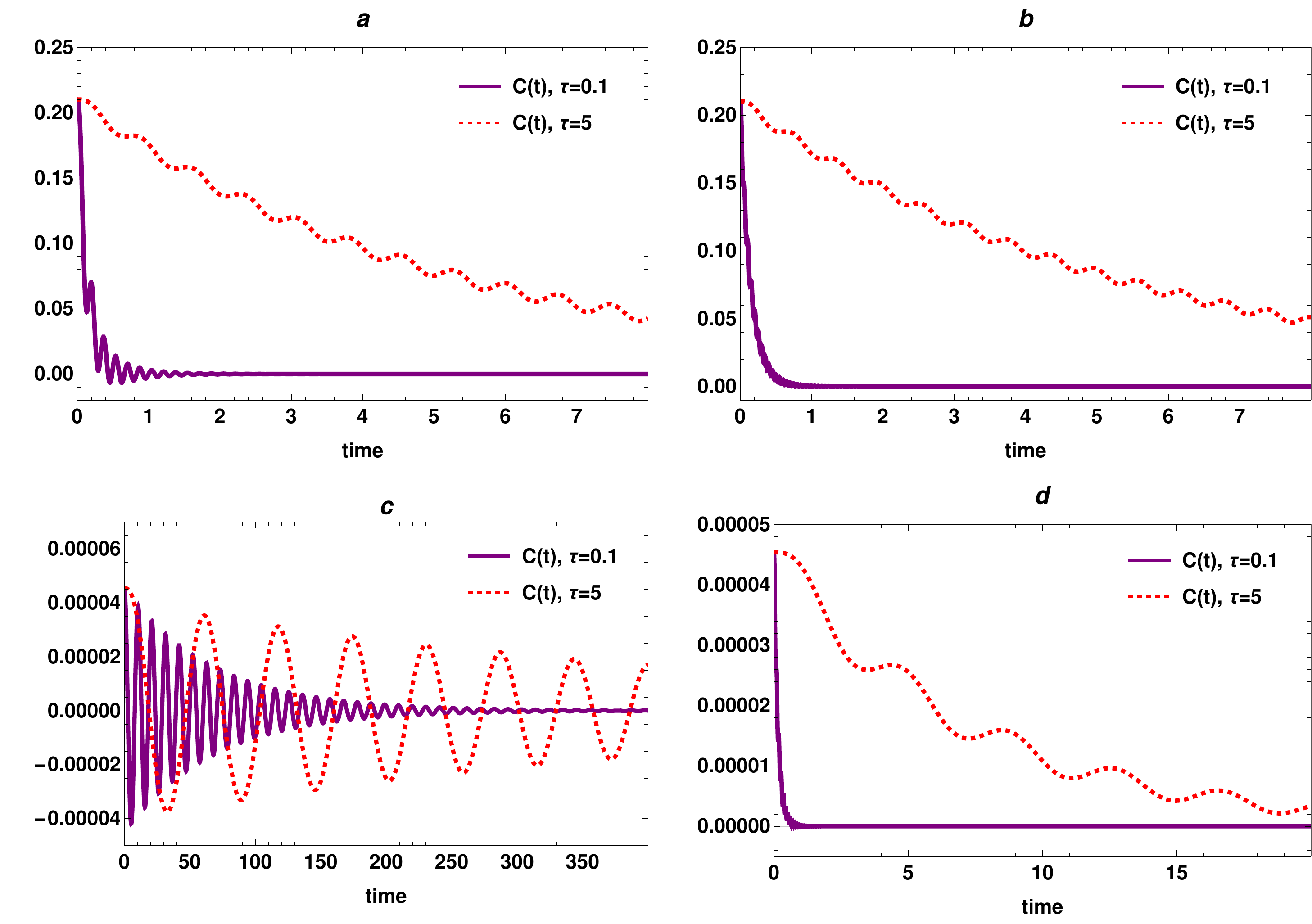}
     \caption{The spin auto-correlation functions ($\mu_B^2$) for the Drude model (memory times $\tau$=0.1 and 5) is plotted as a function of time for the under-damped (low $\gamma$) and the over-damped regimes (high $\gamma$) corresponding to the classical high temperature domain and the quantum low temperature domain with S=1/2, g=1 and $\Omega=10^6$: \textbf{[a]} T=10, $H_0 = 8 $, $\gamma = 5 $, \textbf{[b]} T=10, $H_0=8 $, $\gamma=20 $, \textbf{[c]} T=0.01, $H_0=0.1 $, $\gamma=0.05 $  and  \textbf{[d]} T=0.01, $H_0=0.1 $, $\gamma=5 $.}
     \label{fig5}
 \end{figure}
 
  \begin{figure}[H]
    \centering
    \includegraphics[scale=0.64]{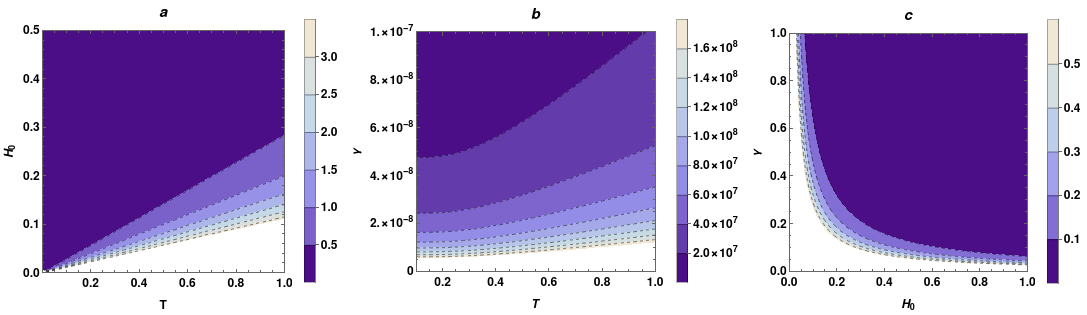}
     \caption{The regions of different relaxation times $\tau_{R}$ for the Ohmic model (S=1/2, g=1 and $\Omega=10^6$) are shown as a function of: \textbf{[a]} T and H$_0$ with $\gamma=0.01 $; \textbf{[b]} T and $\gamma$ with H$_0=1 $; \textbf{[c]} H$_0$ and $\gamma$ with T=0.1. $\tau_{R}$ is 
     indicated in the colour panel on the right.}
     \label{contour_plot}
 \end{figure}
 \end{widetext}
\section{Conclusion}
In this paper we have derived and analysed the quantum Langevin dynamics of a spin in a magnetic field and studied the dynamics of the expectation values of various spin components, the spin auto-correlation functions and the spin response 
function within a fully non-perturbative approach. We have studied these in both the high temperature classical and low 
temperature quantum domains and the results are analysed for an under-damped system and an over-damped system. We have used 
two bath models for our analysis- the memory free Ohmic model and the Drude model with a non-trivial memory and compared the two cases.

Let us summarise our main findings. We have noticed a damped oscillatory behaviour of the expectation values of the spin
components and the spin auto-correlation functions at high and low temperatures at low values of 
the damping coefficient. At higher values of the damping coefficient the damped oscillatory 
behaviour gives way to a monotonically damped behaviour. 

At lower temperature domains, dominated by quantum 
effects, the damping rates are smaller for both under-damped and over-damped regimes for the Drude model due to the presence of a finite memory time.  The larger the value of the Drude memory time the slower the fall off of the various functions. 
This is to 
be contrasted with the memory free Ohmic model, where we find a steeper fall off of the spin components at low temperatures and high damping rates. In addition, we have derived an explicit expression for the spin response function for the Ohmic model which rises from zero to a large peak value and falls off to zero at large times
for a large value of the damping coefficient and exhibits oscillatory damped behaviour for a small value of the damping coefficient. \\
Moreover, we have seen that we recover the results obtained in a semi-classical 
perturbative approach \cite{Dudarev2011} in the limit of a small damping coefficient $\gamma$ (See Eq.(\ref{corrclassical}) ).\\ 
These predictions can be tested in future foultra cold atom systems involving a spin in 
a magnetic field. In the high temperature domain the damped oscillatory behaviour of the spin auto-correlation 
function obtained in our theoretical analysis has indeed been seen in existing cold atom 
experiments involving a spin in a magnetic field \cite{maheswar,spinreview}.
\vspace{0.3cm}

\section*{Acknowledgement}
We acknowledge fruitful discussions with Sanjukta Roy and Saptarishi Chaudhuri regarding the experimental relevance of the work. KM 
acknowledges the hospitality during an academic visit at the Raman Research Institute, Bangalore, where this work has been carried out. 
 \section*{Appendix}
  The component of the spin magnetic moment along the direction of the magnetic field is given by \cite{kittel1996introduction}:
  \begin{align}
    M_z=-gSB_s(x)
\end{align}
where, $x=\frac{SH_0}{k_{B}T}$ and $B_s(x)$ is called the Brillouin function given by \cite{kittel1996introduction,darby1967tables}:
\begin{align}
    B_s(x)=\frac{2S+1}{2S}coth \left[ \left(\frac{2S+1}{2S} \right) x \right]- \frac{1}{2S}coth \left[ \left(\frac{1}{2S} \right) x \right] \label{Brillouinfunc}
\end{align}
In the classical regime, when T is very large, $x<<1$, then
\begin{align}
   & cothx \approx \frac{1}{x}+\frac{x}{3} \notag \\
    & M_z=-gS\left(\frac{x(S+1)}{3S} \right)
\end{align}
In the quantum domain, $T \rightarrow0$, $x>>1$, 
\begin{align}
   & cothx \approx 1 \notag \\
    & M_z=-gS 
\end{align}

Moreover, in the classical limit of large S we get,
\begin{align}
    &coth \left[\frac{x}{2S} \right] \rightarrow \frac{2S}{x}\\
   &  coth \left[\left(1+ \frac{1}{2S}  \right)x \right]  \rightarrow coth (x)
\end{align}
In this limit, $B_s(x)$ goes to,
\begin{align}
  B_s(x)=  coth(x)-\frac{1}{x}
\end{align}
which is known as Langevin function (L(x)) in the Classical limit \cite{darby1967tables}.


%
 \end{document}